\documentclass[journal=jacsat,manuscript=communication,etalmode=truncate,keywords=true]{achemso}
\setkeys{acs}{articletitle=true,maxauthors=10,etalmode=truncate,keywords=true}
\usepackage{amsmath,amssymb,units,txfonts,upgreek}
\usepackage{dblfloatfix}
\usepackage{graphicx}
\usepackage[usenames,dvipsnames]{color}
\usepackage{makecell}

\definecolor{StyleColor}{RGB}{34,80,169} 
\definecolor{abstractcolor}{RGB}{255,243,201}
\usepackage[bookmarks=true,colorlinks=true,urlcolor=StyleColor,linkcolor=StyleColor,citecolor=StyleColor]{hyperref} 
\usepackage{colortbl,cuted,bm}
\usepackage{multirow}
\usepackage{xr}

\newcommand{\eF}{\varepsilon_{\textrm{F}}}

\newcommand{\eVBM}{\varepsilon_{\textrm{VBM}}}
\newcommand{\eCBM}{\varepsilon_{\textrm{CBM}}}
\newcommand{\Hmat}{\mathcal{H}}
\newcommand{\Smat}{\mathcal{S}}

\newcommand{\Tmat}{\mathcal{T}}
\newcommand{\Scalpha}{S_{\!C\alpha}}
\newcommand{\Sc}{S_{\!C}}
\newcommand{\Salpha}{S_{\!\alpha}}
\newcommand{\rv}{\textbf{r}}
\usepackage{xfrac}

\DeclareMathOperator{\Tr}{Tr}

\makeatletter\newenvironment{abstractbox}{%
   \begin{lrbox}{\@tempboxa}\begin{minipage}{0.988\textwidth}}{\end{minipage}\end{lrbox}%
   \colorbox{abstractcolor}{\usebox{\@tempboxa}}
}\makeatother

\renewcommand*\authorfont{\flushleft}
\renewcommand*\affilfont{\mdseries\flushleft\textrm}
\renewcommand*\titlesize{\Large}
\renewcommand*\titlefont{\flushleft\sf\bfseries}
\def\refname{\sffamily\color{StyleColor}
REFERENCES}
\usepackage{titlesec}
\titleformat{\section}{\bfseries\sffamily\color{StyleColor}}{\thesection.~}{0pt}{}
\titleformat{\subsection}[runin]{\bfseries\sffamily\normalsize}{\indent\thesubsection.~}{0pt}{}[.]
\titlespacing{\subsection}{0pt}{0pt}{*1}
\titleformat{\subsubsection}{\bfseries\sffamily\normalsize}{\thethesubsection.~}{0pt}{}
\titlespacing{\subsubsection}{0pt}{0pt}{*0}

\title{Theoretical Insight into the Internal Quantum Efficiencies of Polymer/C$_{\text{60}}$ and Polymer/SWNT Photovoltaic Devices}
\author{Livia No\"{e}mi Glanzmann}
\affiliation[UPV/EHU]{\newline Nano-Bio Spectroscopy Group and ETSF Scientific Development Center, Departamento de F{\'{\i}}sica de Materiales,  Universidad del Pa{\'{\i}}s Vasco UPV/EHU and DIPC, E-20018 San Sebasti\'{a}n, Spain}
\author{Duncan John Mowbray}
\email{duncan.mowbray@gmail.com}
\affiliation[UPV/EHU]{\newline Nano-Bio Spectroscopy Group and ETSF Scientific Development Center, Departamento de F{\'{\i}}sica de Materiales, Universidad del Pa{\'{\i}}s Vasco UPV/EHU and DIPC, E-20018 San Sebasti\'{a}n, Spain}

\begin{document}
\maketitle

\begin{strip}
\vspace{-1.cm}

\noindent{\color{StyleColor}{\rule{\textwidth}{0.5pt}}}
\begin{abstractbox}
\begin{tabular*}{17cm}{b{8.9cm}l}
\noindent\textbf{\color{StyleColor}{ABSTRACT:}}
The internal quantum efficiency (IQE) of an organic photovoltaic device (OPV) is proportional to the number of free charge carriers generated and their conductivity, per absorbed photon.  However, both the IQE and the quantities that determine it, for example, electron--hole binding, charge separation, electron--hole recombination, and conductivity, can only be inferred indirectly from experiments.  Using density functional theory, we calculate the excited-state formation energy,  charge transfer, and zero-bias conductance in the singlet ground state and triplet excited state across polymer/fullerene and polymer/single walled carbon nanotube (SWNT) OPV donor/acceptor bulk heterojunctions.  Specifically, we compare polythiophene (PT) and poly(3-methylthiophene-2,5-diyl) (P3MT) as donors and C$_{60}$ chains with (6,4), (6,5), and (10,5) SWNTs as acceptors.  We find the conduc- &
\includegraphics[height=5cm]{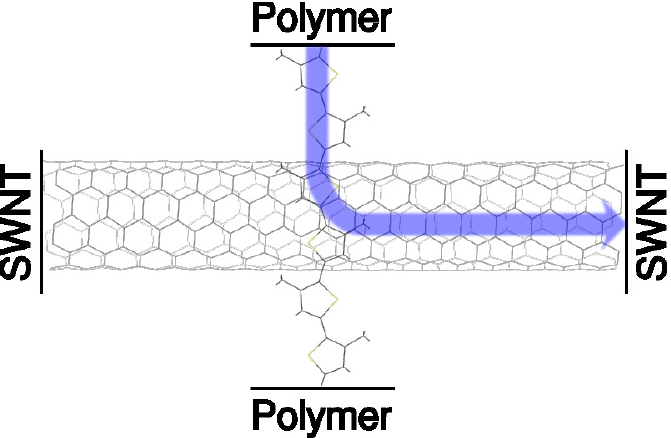}
\\
\multicolumn{2}{p{17cm}}{
tivity  increases substantially for both the excited triplet relative to the singlet ground state and for PT compared with P3MT due to the increased charge transfer and the resulting improvement in donor/acceptor level alignment.  Similarly, the (6,4) SWNT, with a larger SWNT band gap and greater conductivity than fullerenes, provides the highest conductivities of 5 and 9\% of the theoretical maximum for electron and hole carriers, respectively.  This work has important implications for both the optimization of polymer/SWNT bulk heterojunctions and the design of new OPV bulk heterojunctions \emph{in silico}.
}
%
\end{tabular*}
\end{abstractbox}
\noindent{\color{StyleColor}{\rule{\textwidth}{0.5pt}}}
\end{strip}

\section{INTRODUCTION}\label{Sect:Introduction}

The internal quantum efficiency (IQE) of a type II bulk heterojunction within an organic photovoltaic (OPV) device is simply the ratio of free charge carrier generation to photon absorption at a given photon energy.  As such, the IQE depends on the ease of separating electron and holes, and the resulting current through the OPV.  The former depends on the electron--hole binding, charge transfer from donor to acceptor, and electron--hole recombination, while the latter depends on the conductance from donor to acceptor across the bulk heterojunction.  Note that unlike the photovoltaic efficiency,\cite{PhotovoltaicEfficiency} the IQE is independent of the device's optical absorption spectra, that is, the number of absorbed photons, as it is calculated per absorbed photon.\cite{IQE}

To estimate the relative exciton binding between type II bulk heterojunctions, one may compare the first transition for the isolated acceptor, for example, the $E_{11}$ transition of a nanotube, with the formation energy $E_f(\uparrow\uparrow)$ of the triplet excited state from the singlet ground state, that is, their difference in total energy.  A much smaller triplet formation energy for the isolated acceptor, $E_f(\uparrow\uparrow) \ll E_{11}$, suggests that electron--hole separation may prove difficult.  This is the case for single-walled carbon nanotubes (SWNTs), where the measured singlet exciton binding is $\sim 0.4$~eV\cite{AndoJPSJ1997,AvourisPRL2004,SpataruPRL2004,HeinzScience2005,LienauPRB2005}.  Conversely, if the triplet formation energy in the bulk heterojunction is smaller than that for the isolated acceptor, this is indicative of hole transfer to the donor for type II bulk heterojunctions.  In this way, the triplet formation energy can be used to compare the relative ease or difficulty of electron hole separation and the probability of recombination in bulk heterojunctions.  

The ease of electron--hole separation in the bulk heterojunction may also be directly probed by considering the difference in charge transfer from donor to acceptor between the triplet excited state and the singlet ground state\cite{DFT-P3HT-Fullerene,DFT-P3HT-Tubes}.  However, although a greater charge transfer in the triplet excited state means the electron and hole are separated onto different molecules in the bulk heterojunction, it does not address whether the separated electron and hole are free charge carriers or remain bound within the junction. 

To determine whether separated electrons and holes truly behave as free charge carriers, the degree of scattering, and the resulting current through the OPV, one should compute the conductance across the bulk heterojunction from the donor to the acceptor in the excited state.  In particular, one should consider the conductance at the Fermi level of the excited electron/hole to determine the number of free electron/hole carriers.  

In this study, we carry out density functional theory (DFT) calculations of the excited-state formation energy, charge transfer, and zero-bias conductance of prototypical donor/acceptor bulk heterojunctions\cite{Holt,Strano,BHJ2,Ferguson,Kymakis02,Kymakis03,HertelPumpProbeSWNT+PFO-BPy,Bindl,BindlJPCL,Bindl-SWNT,P3HTCNTExpholetransferNanoLett2013,P3HTgapSWNTquenchingJPCC2012,Loi,Lanzi2008Polymer} in the singlet ground state and triplet excited state.  For the donor molecule we employ the prototypical photoactive polymers: polythiophene (PT)\cite{PTgapPRB1987} and poly(3-methylthiophene-2,5-diyl) (P3MT)\cite{P3MTPL}.  For the acceptor molecule we compare fullerene (C$_{60}$) chains with semiconducting (6,4), (6,5), and (10,5) SWNTs, whose band gaps range between 1 and 1.4 eV\cite{Bachilo02S}.  These systems exhibit only a minor structural relaxation in the triplet excited state (<25 meV). This justifies our neglect of molecular vibration as a first approximation,\cite{VibrationsNEGF} as done in previous studies of photovoltaic efficiencies.\cite{ModelingDyeCellsJACS2011,ModelingDyeCellsACR2012,ModelingDyeCellsJPCL2013}

\begin{figure*}
\includegraphics[width=\textwidth]{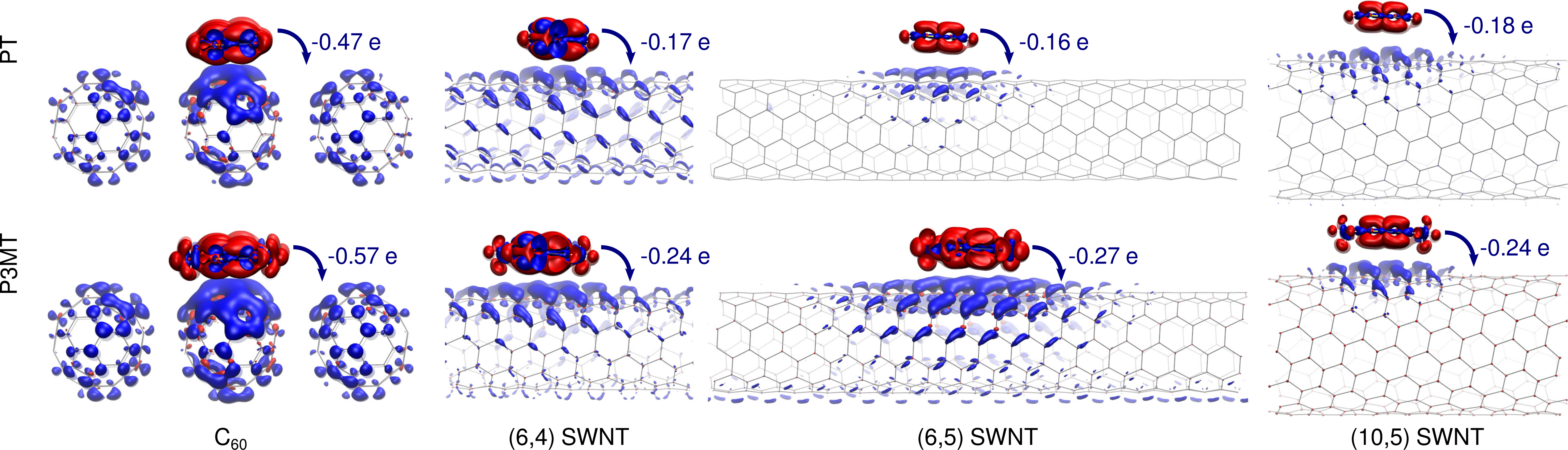}
\caption{Electron (blue) and hole (red) densities at isosurface values of $\pm 1 e/$nm$^3$ from donor (PT or P3MT) to acceptor (C$_{60}$ chain, (6,4), (6,5), or (10,5) SWNTs) from the DFT total charge density difference $\rho(\textbf{r})$ between the excited triplet ($\uparrow\uparrow$) and singlet ($\uparrow\downarrow$) ground state.  
}\label{fgr1}
\noindent{\color{StyleColor}{\rule{\textwidth}{1pt}}}
\end{figure*}

We model the excited-state formation energy using the total energy difference between the triplet excited state and the singlet ground state from DFT.  To obtain the charge transfer from donor to acceptor in the bulk heterojunction, we perform a simplified Bader analysis\cite{Bader} of the DFT all-electron density. Finally, we calculate the Landauer--B\"{u}tticker conductance of electron and hole charge carriers from donor to acceptor across the bulk heterojunction based on the DFT tight-binding Hamiltonians using a multiterminal implementation\cite{cnt_networks} of the nonequilibrium Green's function (NEGF) method\cite{Meir,Datta,KristianWannierFunctions,Thygesen2}.   In this way, we obtain the conductance from the isolated polymer, through the polymer/SWNT or polymer/C$_{60}$ heterojunction, and out the SWNT or C$_{60}$ chain, in the quantum coherent regime.  Our use of the NEGF method for calculating the conductance across polymer/SWNT and polymer/C$_{60}$ heterojunctions is justified by the ballistic transport, long coherence length, and high thermal conductivity of SWNTs\cite{Harris,Dresselhaus,cnt_networks,JuanmaPRL,Juanma}, and the dominance of tunneling in transport processes across C$_{60}$ chains.  This method provides a more sophisticated description of the transport processes, and IQE of the device, than previous studies employing the Newns-Anderson model\cite{ModelingDyeCellsJACS2011,ModelingDyeCellsACR2012,ModelingDyeCellsJPCL2013}.

\section{METHODOLOGY}\label{Sect:Methodology}

All DFT calculations were performed with locally centered atomic orbitals (LCAOs) and the projector augmented wave (PAW) implementation of the \textsc{gpaw} code \cite{GPAW,GPAWRev,GPAWLCAO}, within the generalized gradient approximation (PBE)\cite{PBE} for the exchange correlation (xc)-functional. We employed a double-zeta polarized (DZP) LCAO basis set for representing the density, wave functions, and tight-binding Hamiltonian, which yields transmission functions in quantitative agreement ($\Delta < 50$~meV) with plane-wave codes and maximally localized Wannier functions \cite{BenchmarkPaper}.  All calculations employed a room temperature Fermi filling ($k_B T \approx 25$~meV), with total energies extrapolated to $T\rightarrow 0$~K, that is, excluding the electronic entropy contribution to the free energy $-ST$.  In this way we avoided an unrealistic smearing of the excited electron and hole in the triplet excited state calculations.  We included two thirds of the number of electrons (\sfrac{2}{3}$N_e$) many bands within the calculations. This has been shown to be sufficient to converge the first $\pi\rightarrow\pi^*$ transitions of graphene \cite{DuncanGrapheneTDDFTRPA} and SWNT/polymer hybrid systems \cite{LiviaMilan}, and the optical spectra of polymers and oligomers\cite{Livia2014PSSB} in the random phase approximation (RPA).

Structural optimization was performed within the atomic simulation environment (ASE) \cite{ASE}, until a maximum force below 0.05~eV/\AA{} was obtained. We employed more than 5~\AA{} of vacuum to the cell boundaries orthogonal to the C$_{60}$ chain, (6,4), (6,5), (10,5) SWNTs, polythiophene (PT), poly(3-methylthiophene-2,5-diyl) (P3MT), and poly(3-hexylthiophene-2,5-diyl) (P3HT), and obtained the optimized unit cell parameters along their axes $L_\| =$ 18.811, 40.915, 11.348, 7.867, 7.846, and 7.797~\AA{}, respectively.  Here PT, P3MT, and P3HT are modeled using two thiophene, 3-methylthiophene, and 3-hexylthiophene units, respectively, in $S$-trans configuration.  To sample the Brillouin zone, we included three $k$ points along the axis of PT, P3MT, P3HT, (6,4) SWNT, and (10,5) SWNT and one $k$-point along the axis of the (6,5) SWNT.

The C$_{60}$/polymer junctions were modeled by aligning 10 thiophene/3-methylthiophene units orthogonal to a chain of three C$_{60}$ molecules and fully relaxing the resulting structure, shown in Figure~\ref{fgr1}. The SWNT/polymer junctions were modeled by aligning six thiophene/3-methylthiophene units for the smaller (6,4) and (6,5) SWNTs and eight thiophene/3-methylthiophene units for the (10,5) SWNT orthogonal to the tube, which was repeated once along its axis, and performing single-point calculations for the resulting configurations shown in Figure~\ref{fgr1}.  This repetition of the thiophene/3-methylthiophene units makes a single $k$-point sampling along the polymer axis sufficient for describing the Brillouin zone.

By orienting the polymer and SWNT orthogonal to each other, this configuration describes the limit of a minimal SWNT--polymer overlap.  In fact, by orienting diagonally across the SWNT axis a ten thiophene unit oligomer, which should provide a reasonable description of PT\cite{Livia2014PSSB}, one obtains a greater charge transfer and stronger hybridization between the polymer and SWNT\cite{LiviaMilan}.  

For the SWNT/polymer junctions, the intermolecular distances were fixed between one carbon atom located at the centered C--C single bond of the polymer and a carbon atom of the tube, which were both aligned in the axis orthogonal to the tube and the polymer axis as shown in Figure~\ref{fgr1}. The alignment was achieved by shifting the polymer along the tube axis. This chosen intermolecular C--C distance is 3.35 for PT and 3.39~\AA{} for P3MT.  We slightly increased the P3MT intermolecular distance to reduce the repulsive forces from the hydrogen atoms of the methyl group.  Both minimum distances employed are consistent with the interlayer distance of multiwalled carbon nanotubes (MWNTs) and graphite \cite{graphite-layer}, of $\sim 3.4$~\AA{}.

We performed DFT calculations for each system in both the singlet ground state ($\uparrow\downarrow$) and triplet excited state ($\uparrow\uparrow$).  The triplet excited state calculations were performed by fixing the total magnetic moment $\mu$ of the system, and using separate Fermi levels for the spin majority and minority  channels, $\eF^\uparrow$ and $\eF^\downarrow$, respectively.  The Fermi levels associated with electron $\eF^{e}$ and hole $\eF^h$ charge carriers are then simply $\eF^\uparrow$ and $\eF^\downarrow$, respectively.  For the singlet ground state calculations, $\eF^{e}$ is approximately the conduction band minimum (CBM)  $\eCBM$, while $\eF^h$ is approximately the valence band maximum (VBM) $\eVBM$.

The charge densities associated with the excited electron $\rho_e(\rv)$ and hole $\rho_h(\rv)$ in the triplet state are the negative and positive regions, respectively, of the all-electron charge density difference between that of the triplet excited state ($\uparrow\uparrow$) and singlet ground state ($\uparrow\downarrow$), that is, $\Delta\rho(\rv) = \rho_{\uparrow\uparrow}(\rv) - \rho_{\uparrow\downarrow}(\rv)$.  

To quantify the charge transfer from donor to acceptor, we perform a simplified Bader analysis\cite{Bader} of the all-electron charge density in the singlet ground state $\rho_{\uparrow\downarrow}(\rv)$ and triplet excited state $\rho_{\uparrow\uparrow}(\rv)$.  We begin by integrating the charge density over the plane $A$ of the donor and acceptor axes to obtain the linear charge density $\lambda(z) = \iint_A \rho(r, \varphi, z) r d\varphi dr$.  We then partition $\lambda(z)$ at its minimum $z_{\min}$ in between the polymer and C$_{60}$ chain or SWNT.  The charge transfer from donor to acceptor is then $Q = \int_0^{z_{\min}} \lambda(z)dz + e N_e^{\textrm{acc}}$, where $N_e^{\textrm{acc}}$ is the total number of electrons on the isolated acceptor molecule.  

We employ a similar partitioning to assign a Kohn-Sham (KS) orbital $\psi(\rv)$ to the donor or acceptor molecule of the bulk heterojunction.  The fraction of the $n$th KS orbital $\psi_n(\rv)$ on the acceptor molecule is then $\int_0^{z_{\min}}\iint_A |\psi_n(r,\varphi,z)|^2 r d\varphi dr dz$.   

The periodic DFT Hamiltonian and overlap matrices in the LCAO basis, $\Hmat$ and $\Smat$,  are employed within the NEGF formalism to calculate the Landauer--B\"{u}tticker  conductance for a multiterminal configuration\cite{Datta,Thygesen2,cnt_networks}.  To do so, one must first remove overlap elements between atomic sites separated by more than half the length of the unit cell along the transmission direction, that is, $L_\|/2$, to obtain nonperiodic Hamiltonian and overlap matrices $H$ and $S$.  

The coupling matrix to the semi-infinite leads $V$ for the SWNTs is obtained by repeating the periodic DFT Hamiltonian matrix for the isolated SWNT $\Hmat_{2\times2} = \left(\begin{smallmatrix} H &  \left(\Hmat - H\right)^\dagger\\  \Hmat - H & H \end{smallmatrix}\right)$  
and removing overlap elements between atomic sites separated by more than $L_\|$, yielding the nonperiodic Hamiltonian for semi-infinite lead $\alpha$ $H_\alpha = \left(\begin{smallmatrix} H & V\\ V^\dagger & H\end{smallmatrix}\right)$. 
 This is reasonable for the SWNTs considered herein, for which $L_\| \gtrsim 20$~\AA{}
.  In this way we avoid performing a DFT calculation with a repeated unit cell to obtain the Hamiltonian of the principle layer, $H$, which only couples to the next principle layer through the coupling matrix $V$.   We then align the semi-infinite lead Hamiltonian $H_\alpha$ to the nonperiodic DFT Hamiltonian for the bulk heterojunction using the first onsite energy of the same SWNT carbon atom at the cell boundary.  This is unnecessary for PT, P3MT, and C$_{60}$ because their lead Hamiltonians are extracted directly from the bulk heterojunction Hamiltonian.  In these cases two thiophene units, two 3-methylthiophene units, and a C$_{60}$ molecule comprise a principle layer, with couplings to the next nearest layer numerically zero.

The Hamiltonian matrix for the bulk heterojunction central region, $H_C$, is then generated by augmenting the nonperiodic DFT Hamiltonian for the junction with  the principle layer Hamiltonians $H$ and coupling matrices $V$ of the semi-infinite SWNT leads.  The same procedure is employed to obtain the overlap matrix for the bulk heterojunction central region, $\Sc$.  

Following the multiterminal NEGF procedure described in ref~\citenum{cnt_networks}, the zero-bias conductance at the Fermi level $\eF$ is $G = G_0 \Tr[G_C\Gamma_{\textrm{in}}G_C^\dagger\Gamma_{\textrm{out}}]|_{\varepsilon=\eF}$, where $G_0 = 2e^2/h$ is the quantum of conductance, $G_{C}(\varepsilon)$ is the Green's function of the bulk heterojunction central region, and $\Gamma_{\textrm{in}/\textrm{out}}$ is the coupling to the semi-infinite input and output leads.  The Green's function of the bulk heterojunction central region is $G_C(\varepsilon) = \left[(\varepsilon + i\eta)\Sc - H_C - \sum_\alpha \Sigma_\alpha(\varepsilon)\right]^{-1}$, where $\Sigma_\alpha(\varepsilon)$ is the self-energy of lead $\alpha$, and $\eta \approx 1$~meV is the electronic broadening applied to both the central region and the semi-infinite leads. The self-energy of lead $\alpha$ is $\Sigma_\alpha(\varepsilon) = [(\varepsilon+i\eta)\Scalpha - H_{C\alpha}][(\varepsilon+i\eta)\Salpha-H_\alpha]^{-1}[(\varepsilon+i\eta)\Scalpha^\dagger - H_{C\alpha}^\dagger]$, where $H_{C\alpha}$ and $\Scalpha$ are the couplings of lead $\alpha$ to the bulk heterojunction central region in the Hamiltonian and overlap matrices, $H_C$ and $S_{\!C}$, respectively.  Finally, the coupling to the semi-infinite input/output leads is $\Gamma_{\textrm{in}/\textrm{out}} = i (\Sigma_{\textrm{in}/\textrm{out}} - \Sigma_{\textrm{in}/\textrm{out}}^\dagger)$, where $\Sigma_{\textrm{in}/\textrm{out}}(\varepsilon)$ is the self-energy of the input/output lead.   The four-terminal conductance across the bulk heterojunction from donor to acceptor is then obtained when the input lead is that of a semi-infinite polymer, and the output lead is that of a semi-infinite C$_{60}$ chain or SWNT.

The conductance of electron and hole charge carriers is then obtained by simply evaluating the transmission at their associated Fermi levels $\eF^e$ and $\eF^h$, respectively.  However, to understand the origin of differences in conductivity between the various donor/acceptor bulk heterojunctions considered herein, we shall find it useful to also consider the transmission within an energy window around the acceptor's VBM and CBM.  In this way, we may differentiate between differences in conductivity owing to an improved level alignment between donor and acceptor, an increased overlap of their levels, or a greater number of available transmission channels. 

\section{RESULTS AND DISCUSSION}\label{ResultsandDiscussion}

To quantify the differences in electronic properties among the donors (PT, P3MT, and P3HT) and acceptors (C$_{60}$ chain, (6,4), (6,5), and (10,5) SWNTs), we first consider the energy gaps for the isolated polymers and the band gaps of the isolated fullerene chain and SWNTs.  Among the three functionalized thiophene polymers considered, both the calculated Kohn-Sham (KS) energy gaps and those measured via photoluminesence\cite{PTgapPRB1987,P3MTPL,P3HTgapSWNTquenchingJPCC2012} differ by less than 50 meV, with the largest difference from polythophene (PT) occurring with the first methyl functionalization group (P3MT).  For this reason, along with the accompanying reduction of computational effort, we shall restrict consideration from hereon to PT and P3MT as donors. 

The acceptors considered have been intentionally chosen to provide a range of electronic band gaps, as shown in Table~\ref{TableEgaps}, 
\begin{table}
\vspace{-15.5mm}
\noindent{\color{StyleColor}{\rule{\columnwidth}{1.0pt}}}
\vspace{12.mm}
\caption{\textrm{\bf Kohn-Sham Band Gaps \textit{E}$_\textrm{gap}^\textrm{KS}$, Photoluminescence \textit{E}$_\textrm{11}$ Transitions, and Triplet Excited State Formation Energy \textit{E}$_\textit{f}$($\boldsymbol{\uparrow\uparrow}$) in Electronvolts of Acceptors (C$_{\bf 60}$ chain, (6,4), (6,5), and (10,5) SWNTs) Isolated (---) and in Heterojunctions with a Donor (PT or P3MT)}}\label{TableEgaps}
\begin{tabular}{l|cr@{}lccc}
\multicolumn{1}{>{\columncolor[gray]{.9}}c}{species} & 
\multicolumn{1}{>{\columncolor[gray]{.9}}c}{$E_{\textrm{gap}}^{\textrm{KS}}$}  &
\multicolumn{2}{>{\columncolor[gray]{.9}}c}{$E_{\textrm{11}}$}&
\multicolumn{3}{>{\columncolor[gray]{.9}}c}{$E_f(\uparrow\uparrow)$}
\\[0.5mm]\cline{5-7}
\diaghead{(10,5) SWNT}{acceptor}{donor}& --- & \multicolumn{2}{c}{---} & --- &PT & P3MT\\\hline
C$_{60}$ chain& 1.46 & 1&.9$^a$ 
& 1.508 & 0.842 & 0.622 \\
  (6,4) SWNT & 1.08 & 1&.420$^b$ 
& 1.101 & 0.967 & 0.843\\
  (6,5) SWNT & 0.92 & 1&.272$^b$ 
& 0.853 & 0.804 & 0.689\\
(10,5) SWNT & 0.74 & 0&.992$^b$ 
& 0.858 & 0.712 & 0.626\\
\multicolumn{6}{p{0.7\columnwidth}}{\footnotesize$^a$
Ref~\citenum{C60gapPRB1992}\nocite{C60gapPRB1992}. $^b$Ref~\citenum{Bachilo02S}\nocite{Bachilo02S}.}
\end{tabular}
\noindent{\color{StyleColor}{\rule{\columnwidth}{1.0pt}}}
\end{table}
from 1 eV for the (10,5) SWNT to 1.9 eV for the C$_{60}$ chain.  In each case, the calculated KS band gaps differ from the first $E_{11}$ transition measured in photoluminescence (PL) experiments\cite{C60gapPRB1992,Bachilo02S} by $\sim 25$\%.  This is consistent with previous results, and may be addressed through many-body $GW$ corrections to the self energy\cite{GWTBgraphene,GWSWNT,DarkFullerene}.  However, for our purposes, it is more relevant to note that the difference is rather systematic among the carbon-based materials considered herein.

A more reliable descriptor for differences in exciton binding between bulk heterojunctions is the triplet excited-state formation energy $E_f(\uparrow\uparrow)$, provided in Table~\ref{TableEgaps}.  This is obtained from the total energy difference between a system in the triplet excited state and the singlet ground state, that is, $E_f(\uparrow\uparrow) = E(\uparrow\uparrow) - E(\uparrow\downarrow)$.  Because $E_f(\uparrow\uparrow)$ depends only on DFT total energies and not on KS eigenvalues, this quantity is in principle exact up to the approximation for the xc-functional.  

Comparing the measured $E_{11}$ transition energy and the calculated triplet formation energy $E_f(\uparrow\uparrow)$ for the isolated C$_{60}$ chain and (6,5) SWNT, we find that their difference is $\sim 0.4$~eV.  This is consistent with the measured singlet exciton binding energy in SWNTs\cite{AndoJPSJ1997,AvourisPRL2004,SpataruPRL2004,HeinzScience2005,LienauPRB2005}.  Although differences between the excited singlet and triplet states may occur, this demonstrates that the triplet excited state provides a reasonable approximation to the singlet excited state for fullerenes and SWNTs.  For the (6,4) SWNT, we obtain a somewhat smaller energy difference (0.3~eV), while for the (10,5) SWNT it is significantly reduced to 0.1~eV.  The latter is related to the reduction of the $E_{11}$ transition energy for the (10,5) SWNT, while the $E_f(\uparrow\uparrow)$ is unchanged relative to the (6,5) SWNT.  Altogether, this is indicative of a strong exciton binding in the isolated C$_{60}$ chain, (6,4) SWNT, and (6,5) SWNT, with a much weaker exciton binding in the (10,5) SWNT.  This suggests that charge separation for isolated SWNTs may be easier for those with a smaller band gap.

To compare performance between different acceptors in bulk heterojunctions, we find it is more relevant to compare the difference between the triplet formation energy of the acceptor in the bulk heterojunction and in isolation.  For a type II bulk heterojunction, a lower triplet formation energy in the bulk heterojunction suggests the hole is transferred to the donor.  From Table~\ref{TableEgaps} we see that $E_f(\uparrow\uparrow)$ is reduced for all the acceptors studied when in the bulk heterojunction.  Furthermore, the triplet formation energy is in all cases lower for the P3MT bulk heterojunction than the PT bulk heterojunction.  We also clearly see that the C$_{60}$ chain in the triplet state is significantly more stabilized upon inclusion in the PT or P3MT bulk heterojunctions compared with the SWNTs.  There, $E_f(\uparrow\uparrow)$ is stabilized by $\sim0.1$~eV upon inclusion in a PT bulk heterojunction and a further 0.1 eV for the P3MT bulk heterojunction, for all three SWNTs studied.  Qualitatively, this suggests a significantly greater electron--hole separation for the C$_{60}$ chain compared with the SWNTs, which should all be rather similar, with P3MT bulk heterojunctions having a more facile electron--hole separation than PT bulk heterojunctions.

Alternatively, we may describe the degree of electron--hole separation directly by quantifying the charge transfer in the bulk heterojunction. This is obtained by partitioning the DFT all-electron density between the donor and acceptor species in the bulk heterojunction via a simplified Bader analysis. The charge transfer from donor to acceptor for each bulk heterojunction in the triplet excited state, singlet ground state, and their difference, is provided in Table~\ref{TableChargeTransfer}.
\begin{table}
\vspace{-12mm}
\noindent{\color{StyleColor}{\rule{\columnwidth}{1.0pt}}}
\vspace{8mm}
\caption{\textrm{\bf Charge Transfer in \textit{e} from Donor (PT or P3MT) to Acceptor (C$_{\bf60}$ chain, (6,4), (6,5), or (10,5) SWNTs) for the Singlet Ground State ($\boldsymbol{\uparrow\downarrow}$), the Triplet Excited State ($\boldsymbol{\uparrow\uparrow}$), and Their Difference ($\boldsymbol{\uparrow\uparrow - \uparrow\downarrow}$)}}\label{TableChargeTransfer}
\begin{tabular}{llccc}
\multicolumn{1}{>{\columncolor[gray]{.9}}c}{donor} & 
\multicolumn{1}{>{\columncolor[gray]{.9}}c}{acceptor} & 
\multicolumn{1}{>{\columncolor[gray]{.9}}c}{$\uparrow\downarrow$} & 
\multicolumn{1}{>{\columncolor[gray]{.9}}c}{$\uparrow\uparrow$}&
\multicolumn{1}{>{\columncolor[gray]{.9}}c}{$\uparrow\uparrow-\uparrow\downarrow$}
\\[0.5mm]
PT & C$_{60}$ chain& $-$0.01 & $-$0.48 & $-$0.47\\
P3MT & C$_{60}$ chain&$-$0.04 & $-$0.61 & $-$0.57\\
\hline
PT & (6,4) SWNT &$-$0.05 & $-$0.22 & $-$0.17\\
P3MT & (6,4) SWNT &$-$0.10 & $-$0.34 & $-$0.24\\
\hline
PT & (6,5) SWNT &$-$0.16 & $-$0.31 & $-$0.16\\
P3MT & (6,5) SWNT &$-$0.23 & $-$0.50 & $-$0.27\\
\hline
PT & (10,5) SWNT &$-$0.10 &$-$0.29 & $-$0.18\\
P3MT & (10,5) SWNT &$-$0.19 & $-$0.43 & $-$0.24\\
\end{tabular}
\noindent{\color{StyleColor}{\rule{\columnwidth}{1.0pt}}}
\end{table}

From Table~\ref{TableChargeTransfer} we observe the following trends in the charge transfer: (1) It is always from donor to acceptor, (2) it is always more for P3MT than PT bulk heterojunctions by about $-0.1$~$e$, (3) for SWNTs it is always about $-0.25$~$e$ for P3MT bulk heterojunctions and $-0.17$~$e$ for PT bulk heterojunctions, and (4) it is significantly greater ($-0.5$~$e$) for C$_{60}$ chains.   All four findings are consistent with our previously mentioned expectations based on the triplet excited-state formation energy.  

It is useful to consider the spatial distribution of the difference in all electron density between triplet excited state and singlet ground state shown in Figure~\ref{fgr1} to determine the origin of these trends in the bulk heterojunctions' charge transfer.  For each bulk heterojunction studied, the hole density is mostly localized on the $\pi$-bonding highest occupied molecular orbital (HOMO) of PT or P3MT, with the electron density on $\pi$ antibonding levels of the C$_{60}$ chain or SWNT.  Comparing the PT and P3MT bulk heterojunctions, we notice that the hole density clearly extends onto the methyl groups of P3MT.  We expect this spatial delocalization of the hole density onto the methyl groups of P3MT makes hole transfer easier in P3MT than PT bulk heterojunctions.  For the (6,4), (6,5) and (10,5) SWNT bulk heterojunctions, the electron and hole densities shown in Figure~\ref{fgr1} are rather consistent, with the electron density on the upper surface of the SWNT neighboring the PT or P3MT.  Conversely, for the C$_{60}$ chain, the electron density is delocalized over the entire surface of all three fullerenes.  This suggests the greater charge transfer onto the C$_{60}$ chains compared with the SWNTs may have a geometrical origin.  

However, although these results clearly demonstrate a charge transfer from donor to acceptor, it remains unclear whether the excited electron and hole are truly free charge carriers or remain bound at the donor/acceptor interface.  To address this issue, we provide the zero-bias conductance at the energy of the excited electron $\eF^e$ and hole $\eF^h$ in the singlet ground state and triplet excited state in Table~\ref{TableConductance}.
\begin{table}
\vspace{-15.5mm}
\noindent{\color{StyleColor}{\rule{\columnwidth}{1.0pt}}}
\vspace{10mm}
\caption{\textrm{\bf Conductance \textit{G} in \textit{G}$_\textrm{0}\boldsymbol{\times10}^{\boldsymbol-\boldsymbol3}$ from Donor (PT or P3MT) to Acceptor (C$_{\bf60}$ chain, (6,4), (6,5), or (10,5) SWNTs) for Hole Carriers at $\boldsymbol\varepsilon_\textrm{F}^\textit{h}$ and Electron Carriers at $\boldsymbol\varepsilon_\textrm{F}^\textit{e}$ in the Singlet Ground State ($\boldsymbol{\uparrow\downarrow}$) and the Triplet Excited State ($\boldsymbol{\uparrow\uparrow}$)}}\label{TableConductance}
\begin{tabular}{lll@{\hspace{3.mm}}r@{}ll@{\hspace{3.mm}}r@{}l}
\multicolumn{2}{>{\columncolor[gray]{0.9}}c}{}&
\multicolumn{3}{>{\columncolor[gray]{.9}}c}{\underline{hole carriers}}&
\multicolumn{3}{>{\columncolor[gray]{.9}}c}{\underline{electron carriers}}
\\
\multicolumn{1}{>{\columncolor[gray]{.9}}c}{donor} & 
\multicolumn{1}{>{\columncolor[gray]{.9}}c}{acceptor} &
\multicolumn{1}{>{\columncolor[gray]{.9}}c}{$\uparrow\downarrow$} & 
\multicolumn{2}{>{\columncolor[gray]{.9}}c}{$\uparrow\uparrow$}&
\multicolumn{1}{>{\columncolor[gray]{.9}}c}{$\uparrow\downarrow$}&
\multicolumn{2}{>{\columncolor[gray]{.9}}c}{$\uparrow\uparrow$}
\\[0.5mm]
PT & C$_{60}$ chain&
0.0007 & 0&.003 & 0.003 & 0&.02
\\
P3MT & C$_{60}$ chain&
0.3 & 0&.0007 & 0.09& 0&.6
\\\hline
PT & (6,4) SWNT & 
0.05& 88& & 0.02 & 46&
\\
P3MT & (6,4) SWNT &
0.0005& 1& & 0.01 & 16&
\\\hline
PT & (6,5) SWNT &
0.0007& 0&.007 &0.02 & 0&.005
\\
P3MT & (6,5) SWNT &
0.0002& 0&.01 & 0.0003 &0&.004
\\\hline
PT & (10,5) SWNT &
0.01& 0&.01 & 0.002 & 0&.003
\\
P3MT & (10,5) SWNT &
0.0003 & 0&.04 & 0.0002 &0&.002
\\
\end{tabular}
\noindent{\color{StyleColor}{\rule{\columnwidth}{1.0pt}}}
\end{table}

Note that these conductances are across a single fullerene/polymer or SWNT/polymer junction.  This is because within the NEGF formalism, the fullerene chain, SWNT, and polymer are all modeled as semi-infinite leads.  For this reason, the conductances provided in Table~\ref{TableConductance} may be considered to be per absorbed photon. Furthermore, because there is only a single PT or P3MT band for this energy range through which current may flow, the quantum of conductance, $G_0 = 2 e^2/h$, provides a theoretical upper bound on the conductance across the junction.

\begin{figure*}[!t]
\includegraphics[scale=0.45]{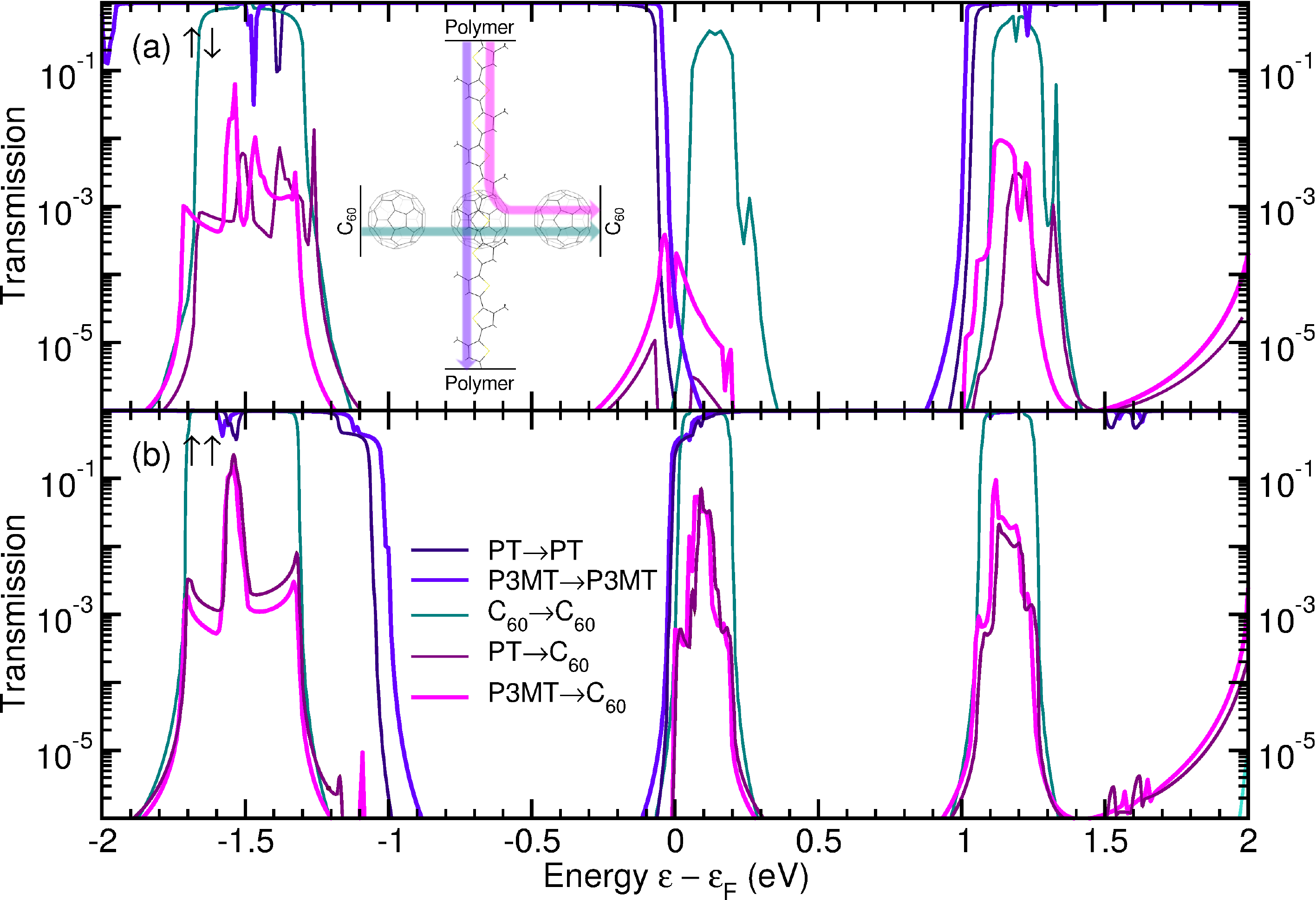}
\caption{Transmission through a polymer/C$_{60}$ junction, depicted schematically as an inset, versus energy, $\varepsilon$, in electronvolts relative to the Fermi level $\eF$ in the (a) singlet ground state ($\uparrow\downarrow$) and (b) triplet excited state ($\uparrow\uparrow$) for PT/C$_{60}$ (thin dark lines) and P3MT/C$_{60}$ (thick light lines).
}\label{fgr2}
\end{figure*}
\begin{figure*}[!t]
\includegraphics[scale=0.45]{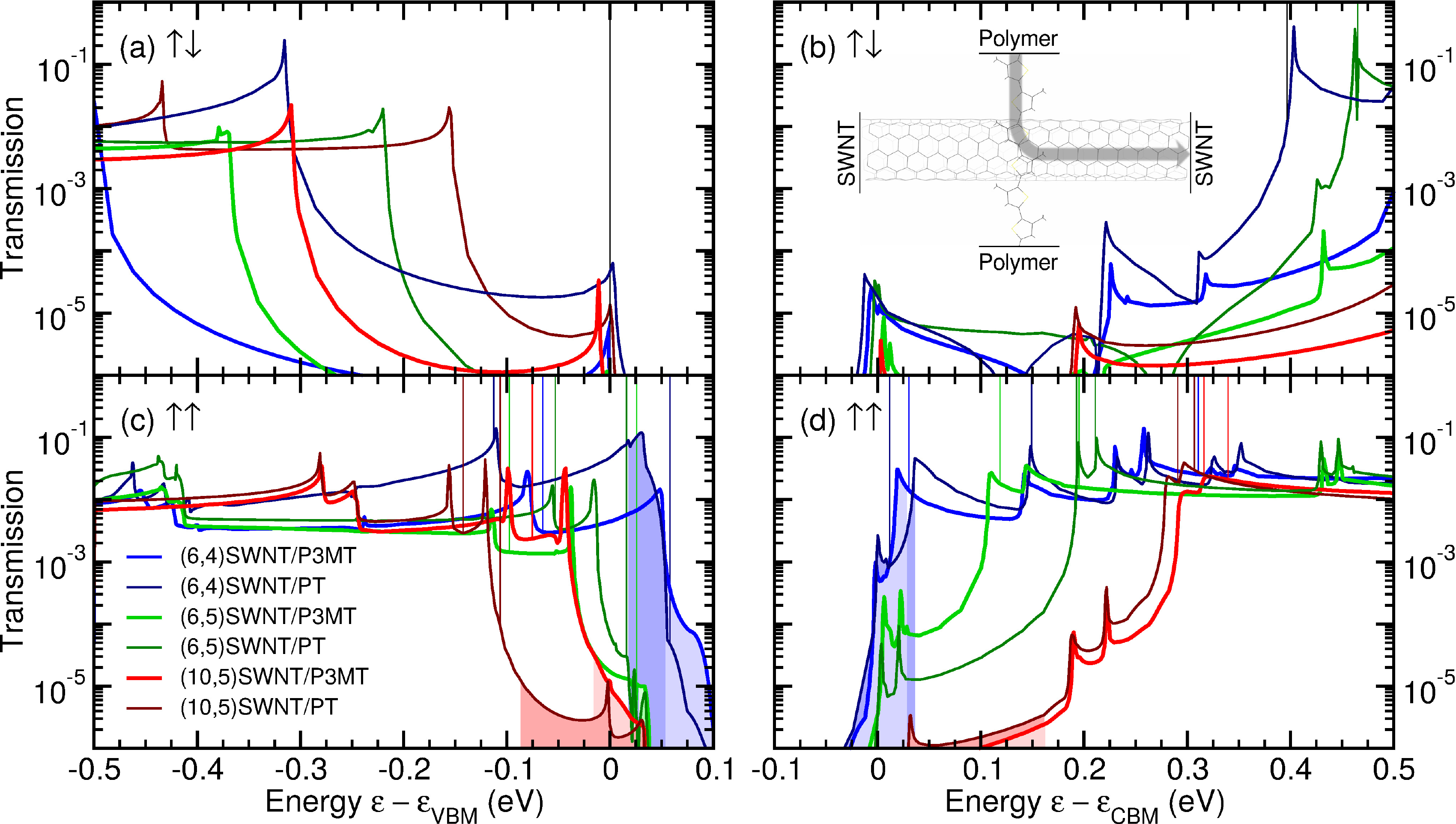}
\caption{Transmission across a polymer/SWNT junction, depicted schematically as an inset, versus energy, $\varepsilon$, in electronvolts of (a,c) the hole relative to the valence band (VB) maximum, $\varepsilon_{\textrm{VBM}}$, and (b,d) the electron relative to the conduction band (CB) minimum, $\varepsilon_{\textrm{CBM}}$, in the (a,b) singlet ground state ($\uparrow\downarrow$) and (c,d) triplet excited state ($\uparrow\uparrow$), with PT (thick light lines) and P3MT (thin dark lines) as donors and a (6,4) (blue), (6,5) (green), and (10,5)SWNT (red) as acceptors.  DFT eigenenergies of the HOMO, SOMO, SUMO, and LUMO of P3MT/PT are marked by thin vertical lines.  Filling in  (c,d) denotes hole/electron charge carriers above/below the Fermi level of the VB/CB in the triplet excited state.
}\label{fgr3}
\noindent{\color{StyleColor}{\rule{\textwidth}{1.0pt}}}
\end{figure*}

Overall, the conductances quoted in Table~\ref{TableConductance} vary by over 6 orders of magnitude, from $G_0\times10^{-7}$ to $0.1 G_0$.  Particularly impressive is the (6,4) SWNT/PT junction, with conductances of 9 and 5\% of $G_0$ for free hole and electron carriers, respectively.  While the (6,4) SWNT junctions clearly provide the highest conductivity in the triplet excited state, the C$_{60}$ chain/P3MT junction is clearly the most active in the singlet ground state.  Comparing the conductance for the junction in the singlet ground state and the triplet excited state, we find that the conductance is generally significantly greater when the system is in the triplet excited state.  This suggests that the level hybridization related to the charge transfer observed in Figure~\ref{fgr1} facilitates tunnelling across the donor/acceptor gap.  However, the opposite is true for the C$_{60}$ chain/P3MT junction.   We also find the (6,5) and (10,5) SWNTs, with their smaller band gaps, have quite low conductivities.  This is consistent with recent experimental findings, which showed that (6,5)SWNT/PCBM/P3HT bulk heterojunctions perform rather poorly\cite{
Loi}.

To provide insight into the reasons behind the great variability in the free carrier conductance provided in Table~\ref{TableConductance} and the origin of the high conductance across the (6,4) SWNT junctions in the triplet excited state, we plot the transmission function $\Tmat(\varepsilon) = \Tr[G_C\Gamma_{\textrm{in}}G_C^\dagger\Gamma_{\textrm{out}}]$ near the VBM and CBM for the C$_{60}$/polymer and SWNT/polymer junctions in Figures~\ref{fgr2} and \ref{fgr3}, respectively.

For the C$_{60}$/polymer junction, depicted schematically as an inset in Figure~\ref{fgr2}(a), we provide the conductance through the polymers (PT$\rightarrow$PT and P3MT$\rightarrow$P3MT), along the C$_{60}$ chain (C$_{60}\rightarrow$C$_{60}$), and across the junction (PT$\rightarrow$C$_{60}$ and P3MT$\rightarrow$C$_{60}$).  The transmission along the polymers is simply $G_0$ for energies below the HOMO, zero within the energy gap, and $G_0$ above the lowest unoccupied molecular orbital (LUMO).  This amounts to a simple counting of the number of bands below the VBM and above the CBM.  

In the singlet ground state, the HOMO of PT/P3MT is pinned to the LUMO of C$_{60}$ at the Fermi level, while in the triplet excited state, the LUMO of PT/P3MT becomes pinned to the LUMO of C$_{60}$ at the Fermi level of the excited electron $\eF^e$.  This means any differences in band gap between PT and P3MT do not play an important role for the polymer/fullerene bulk heterojunctions, as the donor HOMO and LUMO are pinned to the LUMO of the acceptor in the ground and excited states, respectively.  

Conduction through the C$_{60}$ chain is rather different.  It instead exhibits narrow plateaus centered on the HOMO and LUMO of the C$_{60}$ chain.  As a result, conduction across the polymer/fullerene junction is limited to these narrow plateaus where the C$_{60}$ chain is conductive, as shown in Figure~\ref{fgr2}a,b.  It is this limitation in the conductivity of C$_{60}$ chains, which limits their effectiveness within OPV bulk heterojunctions and motivates their replacement with SWNTs\cite{P3HTgapSWNTquenchingJPCC2012}.

Unlike the C$_{60}$ chain, the transmission through semiconducting SWNTs exhibits broad plateaus, and is typically $2G_0$ below the VBM, zero within the band gap, and $2G_0$ above the VBM.  As was this case for transmission through the polymers, this amounts to a simple counting of the number of bands below and above the VBM and CBM, respectively.  Because the transmission through the polymers and SWNTs is rather trivial, up to scattering due to transmission across the junction,  it has been omitted in Figure~\ref{fgr3}.  Instead, we plot the HOMO, singly occupied molecular level (SUMO), singly unoccupied molecular level (SOMO), and LUMO energies below/above which the transmission from PT$\rightarrow$PT and P3MT$\rightarrow$P3MT is $G_0$ in Figure~\ref{fgr3}a--d. 

For the SWNT/polymer junctions, we find the conductance is intimately related to the alignment of the SWNT/polymer HOMO and LUMO levels (\emph{cf}. Figure~\ref{fgr3}a--d).  In fact, the conductance at $\eF^e$ and $\eF^h$ depends exponentially on the alignment of the SWNT and polymer HOMOs and LUMOs \cite{JCPMowbray}.  

For the (6,4) SWNT, with a KS band gap a bit smaller than PT and P3MT, we find both the HOMO and LUMO levels of the polymer and SWNT are aligned in the triplet excited state.  Although the resulting charge transfer is rather similar to the other SWNTs studied (\emph{cf.} Table~\ref{TableChargeTransfer}), the improvement in level alignment places $\eF^h$ and $\eF^e$ at or near the polymer HOMO and LUMO, respectively.  This results in a very high conductivity across the (6,4) SWNT/polymer junctions in the triplet excited state (\emph{cf.}~Table~\ref{TableConductance}).  Conversely, the level alignment for the (10,5) SWNT is the poorest of those considered, as the CBM of the nanotube is shifted down relative to  the LUMO of the polymer as the band gap decreases.  

In fact, if the transmission was measured above or below the polymer LUMO or HOMO, respectively, the conductance across all three polymer/SWNT junctions studied would be $>1$\% of $G_0$.  This is not the case for the C$_{60}$ chain, which is only conductive within a narrow range of the chain's HOMO and LUMO.

Essentially, the conductance across SWNT/polymer heterojunctions can be dramatically improved by a better alignment of the polymer's HOMO and LUMO levels with the SWNT's VBM and CBM, respectively.  This may be accomplished using a SWNT with a larger band gap, for example, a (6,4) SWNT.  However, improving the level alignment too much may lead to a reduction of the electron and hole transfer.  

\section{CONCLUSIONS}\label{Sect:Conclusions}

We have employed four descriptors: (1) the triplet state formation energy $E_f(\uparrow\uparrow)$, (2) the donor to acceptor charge transfer, (3) the conductance of free electron carriers $G(\eF^e)$, and (4) the conductance of free hole carriers $G(\eF^h)$, to assess the relative performance of OPV bulk heterojunctions with PT and P3MT as donors, and C$_{60}$ chains, (6,4), (6,5), and (10,5) SWNTs as acceptors.  We find P3MT, with its larger band gap, and greater ability to absorb a hole, generally exhibits a greater charge transfer and conductance than PT.  The C$_{60}$ chain accepts significantly more charge ($-0.5$~$e$) than the SWNTs ($-0.25$~$e$), which are rather consistent for all SWNTs considered.  These results are also consistent with the calculated formation energies for the triplet excited state.  

However, the conductance across the junctions via free hole and electron carriers differs by six orders of magnitude among the bulk heterojunctions considered here. In the singlet ground state the C$_{60}$ chain/P3MT junction has the greatest free hole and electron conductivity of those considered (0.03 and 0.01\% of $G_0$), while the (6,4) SWNT/PT junction shows a dramatic increase in conductivity in the triplet excited state (9 and 5\%).  This suggests that by improving the level alignment of the polymer and SWNT through the use of larger band gap SWNTs one may obtain a dramatic improvement in OPV efficiency.

Altogether, these results demonstrate the importance of considering the hybridization of donor/acceptor levels in the excited state, and the resulting dependence on level alignment of the conductivity.  This dramatic dependence on the level alignment observed herein provides significant motivation for future studies including the dependence on the vibrational modes of the molecules\cite{VibrationsNEGF}, and more advanced quasiparticle calculations including anisotropic screening effects\cite{OurJACS,MiganiLong,MiganiH2OJCTC2014,SunH2OAnatase,Catechol} in the polymer/SWNT level alignment.  Furthermore, a reformulation of the NEGF method to describe photoinduced quantum transport is required to describe the optical absorption-dependent photovoltaic efficiency of OPV devices.  The techniques employed herein provide a roadmap for the computational design of OPV bulk heterojunctions \emph{in silico}.  


\section*{\large$\blacksquare$\normalsize{} AUTHOR INFORMATION}
\subsubsection*{Corresponding Author}
\noindent*E-mail: \href{mailto:duncan.mowbray@gmail.com}{duncan.mowbray@gmail.com}. Tel: +34 943 01 8392.
\subsubsection*{Notes} 
\noindent The authors declare no competing financial interest.
\section*{\large$\blacksquare$\normalsize{} ACKNOWLEDGMENTS} 

The authors thank Angel Rubio and Llu\'{\i}s Blancafort for fruitful discussions.  We acknowledge financial support from the European Projects POCAONTAS (FP7-PEOPLE-2012-ITN-316633), DYNamo (ERC-2010-AdG-267374), MOSTOPHOS (SEP-210187476),  and EUSpec (COST Action MP1306); Spanish Grants (FIS20113-46159-C3-1-P) and ``Grupos Consolidados UPV/EHU del Gobierno Vasco'' (IT-578-13); the Air Force Office of Scientific Research (AFOSR) (FA2386-15-1-0006 AOARD 144088); and computational time from the BSC Red Espanola de Supercomputacion.


\bibliography{bibliography}


\end{document}